\documentclass[aps,11pt,twoside,nofootinbib]{revtex4}
\pdfoutput=1
\usepackage{amsthm}
\usepackage{amsmath,latexsym,amssymb,verbatim,enumerate,graphicx}
\usepackage{color}

\usepackage{mathtools}
\usepackage{hyperref}

\newtheorem*{rep@theorem}{\rep@title}
\newcommand{\newreptheorem}[2]{%
\newenvironment{rep#1}[1]{%
 \def\rep@title{#2 \ref{##1} (restatement)}%
 \begin{rep@theorem}}%
 {\end{rep@theorem}}}
\makeatother

\newreptheorem{thm}{Theorem}
\newreptheorem{lem}{Lemma}

\def\ba#1\ea{\begin{align}#1\end{align}}
\def\ban#1\ean{\begin{align*}#1\end{align*}}

\newcommand{\ot}{\otimes}
\newcommand{\be}{\begin{equation}}
\newcommand{\ee}{\end{equation}}

\def\benum{\begin{enumerate}}
\def\eenum{\end{enumerate}}

\def\squareforqed{\hbox{\rlap{$\sqcap$}$\sqcup$}}
\def\qed{\ifmmode\squareforqed\else{\unskip\nobreak\hfil
\penalty50\hskip1em\null\nobreak\hfil\squareforqed
\parfillskip=0pt\finalhyphendemerits=0\endgraf}\fi}
\def\endenv{\ifmmode\;\else{\unskip\nobreak\hfil
\penalty50\hskip1em\null\nobreak\hfil\;
\parfillskip=0pt\finalhyphendemerits=0\endgraf}\fi}

\newcommand{\ket}[1]{|#1\rangle}

\newcommand{\tr}{\text{tr}}

\newcommand{\id}{\mathbb{I}}

\newcommand{\<}{\langle}
\renewcommand{\>}{\rangle}

\def\id{{\operatorname{id}}}

\def\be{\begin{equation}}
\def\ee{\end{equation}}
\def\ben{\begin{eqnarray}}
\def\een{\end{eqnarray}}
\def\ot{\otimes}
\def\bei{\begin{itemize}}
\def\eei{\end{itemize}}

\mathchardef\ordinarycolon\mathcode`\:
\mathcode`\:=\string"8000
\def\vcentcolon{\mathrel{\mathop\ordinarycolon}}
\begingroup \catcode`\:=\active
  \lowercase{\endgroup
  \let :\vcentcolon
  }

\newcommand{\nc}{\newcommand}
 \nc{\proj}[1]{|#1\rangle\!\langle #1 |} 
\nc{\avg}[1]{\langle#1\rangle}

\nc{\conv}{\operatorname{conv}}
\nc{\smfrac}[2]{\mbox{$\frac{#1}{#2}$}} \nc{\Tr}{\operatorname{Tr}}
\nc{\ox}{\otimes} \nc{\dg}{\dagger} \nc{\dn}{\downarrow}
\nc{\lmax}{\lambda_{\text{max}}}
\nc{\lmin}{\lambda_{\text{min}}}

\nc{\csupp}{{\operatorname{csupp}}}
\nc{\qsupp}{{\operatorname{qsupp}}} \nc{\var}{\operatorname{var}}
\nc{\rar}{\rightarrow} \nc{\lrar}{\longrightarrow}
\nc{\poly}{\operatorname{poly}}
\nc{\polylog}{\operatorname{polylog}} \nc{\Lip}{\operatorname{Lip}}
\nc{\Om}{\Omega}
\nc{\wt}[1]{\widetilde{#1}}

\def\>{\rangle}
\def\<{\langle}

\nc{\glneq}{{\raisebox{0.6ex}{$>$}  \hspace*{-1.8ex} \raisebox{-0.6ex}{$<$}}}
\nc{\gleq}{{\raisebox{0.6ex}{$\geq$}\hspace*{-1.8ex} \raisebox{-0.6ex}{$\leq$}}}

\nc{\vholder}[1]{\rule{0pt}{#1}}
\nc{\wh}[1]{\widehat{#1}}
\nc{\h}[1]{\widehat{#1}}

\nc{\ob}[1]{#1}

\def\beq{\begin {equation}}
\def\eeq{\end {equation}}

\def\be{\begin{equation}}
\def\ee{\end{equation}}
\def\AA{A}
\def\BB{B}

\nc{\eq}[1]{(\ref{eq:#1})} 
\nc{\eqs}[2]{\eq{#1} and \eq{#2}}

\nc{\eqn}[1]{Eq.~(\ref{eqn:#1})}
\nc{\eqns}[2]{Eqs.~(\ref{eqn:#1}) and (\ref{eqn:#2})}

\newcommand{\temp}{\mathrlap{\otimes }^{\,T}}
\def\id{\mathbb{I}}
\nc{\region}{\cS\cW}

\newenvironment{protocol*}[1]
  {
    \begin{center}
      \hrulefill\\
      \textbf{#1}
  }
  {
    \vspace{-1\baselineskip}
    \hrulefill
    \end{center}
  }

\begin{document}


\title{Do black holes create polyamory?}

\author{Andrzej Grudka$^{1,4}$, Michael J. W. Hall$^2$, Micha\l{} Horodecki$^{3,4}$, Ryszard Horodecki$^{3,4}$, Jonathan Oppenheim$^5$,
John A. Smolin$^6$}
\affiliation{$^1$Faculty of Physics, Adam Mickiewicz University, 61-614 Pozna\'n, Poland }
\affiliation{$^2$Centre for Quantum Computation and Communication Technology (Australian Research Council),
Centre for Quantum Dynamics, Griffith University, Brisbane, QLD 4111, Australia}
\affiliation{$^3$Institute of Theoretical Physics and Astrophysics, University of Gda\'nsk, Gda\'nsk, Poland}
\affiliation{$^4$National Quantum Information Center of Gda\'nsk, 81--824 Sopot, Poland}
\affiliation{$^5$University College of London, Department of Physics \& Astronomy, London, WC1E 6BT and London Interdisciplinary Network for Quantum Science} 
\affiliation{$^6$IBM T. J. Watson Research Center, 1101 Kitchawan Road, Yorktown Heights, NY 10598}


\begin{abstract}
  Of course not, but if one believes that information cannot be destroyed in a theory of quantum gravity, then we run into apparent contradictions with quantum theory when we consider evaporating black holes. Namely that the no-cloning theorem or the principle of entanglement monogamy is violated. Here, we show that neither violation need hold,
since, in arguing that black holes lead to cloning or non-monogamy, 
one needs to assume a tensor product structure between two points in space-time that could instead be viewed as 
causally connected. In the latter case,  one is violating the semi-classical causal structure of space, which is a strictly
weaker implication than cloning or non-monogamy. This is because both cloning and non-monogamy also lead to a breakdown of the semi-classical causal structure.
We show that the lack of monogamy that can emerge in evaporating space times is one that 
is allowed in quantum  mechanics, and is very naturally related to a lack of monogamy of correlations of outputs of measurements performed at subsequent instances of time of a single system. 
This is due to an interesting duality between temporal correlations and entanglement. 
A particular example of this is the Horowitz-Maldacena proposal, and we argue that it needn't lead to cloning or violations of entanglement monogamy.  
For measurements on systems which appear to be leaving a black hole, we introduce the notion of the temporal product, and argue
that it is just as natural a choice for measurements as the tensor product. For black holes, the tensor and temporal products have the same measurement statistics, but result in different type of non-monogamy of correlations, with the former being forbidden in quantum theory while the latter is allowed. In the case of the AMPS firewall experiment we find that the entanglement structure
is modified, and one must have
entanglement between the infalling Hawking partners and early time outgoing Hawking radiation which surprisingly tame the violation of entanglement monogamy.    
\end{abstract}

\maketitle


\section{Introduction: Polyamory and cloning versus temporal products}

In the original black hole information loss problem it is claimed that
unitarity leads to cloning of quantum
states \cite{susskind-thorlacius-note-preskill}, which is not allowed in quantum
theory. In the AMPS {\it gedanken experiment}, or firewall
problem~\cite{almheiri2013black,braunstein2009entangled}, it is
claimed that unitarity leads to a violation of the principle of
entanglement monogamy -- namely that if a system is maximally
entangled with another system, it cannot be correlated with a third
system \cite{coffman2000distributed,bennet-monogomy,koashi2004monogamy}. Again,
this is something which is not allowed in quantum theory, and we say that systems which
violate this property are entangled in a manner which is {\it polyamorous}
\footnote{We will be distinguishing polyamorous entanglement (where many systems can
  be entangled with each other) from polygamous entanglement (where one system is entangled with multiple systems). The latter is associated with the AMPS firewall experiment, while we
find that the former can be present as well}. 

   In this paper we do not aim to solve the black hole information problem.
Rather, we wish to point out that, if we insist on unitary evolution, what we appear to be sacrificing is the standard causal structure of general relativity. However, we do not need to additionally sacrifice violations of quantum theory, such as the no-cloning theorem and monogamy of entanglement.   


The argument that black holes lead to cloning can be generalised
beyond black hole space-times and goes like this: If evolution takes a
state outside of its own light cone, for example, if an unknown state
of some system is replicated at two space-like separated points A and
B, and our theory is relativistically invariant, then there exists a
reference frame in which the state has evolved from an initial single
copy, to two copies of the state, one at A and one at B (see Figure
\ref{fig:cloning}).  Such an evolution taking a single copy of a
state to two copies cannot be unitary\footnote{In some communities,
  the term unitary is sometimes taken to mean probability (or trace)
  preserving. Here, we just mean that pure states evolve to pure
  states via a unitary matrix $U$, $UU^\dagger=\id$.}  or even linear
\cite{WoottersZurek1982}.  In the case of the black hole, one finds a family
of space-like hypersurfaces known as ''nice-slices'' \cite{wald1992space} which are
well-away from the singularity, yet intersect almost all the outgoing
Hawking radiation as well as the in-falling matter which formed or fell into the
black hole.  These hypersurfaces can
contain two copies of the state, the one inside the black hole, and
the one outside.
Thus if information eventually escapes the black hole, it is claimed
that the no-cloning theorem (and hence, linearity), would be violated.

\begin{figure}[h]
\centering
\includegraphics[trim=1cm 0cm 5cm 0cm, totalheight=0.3\textheight]{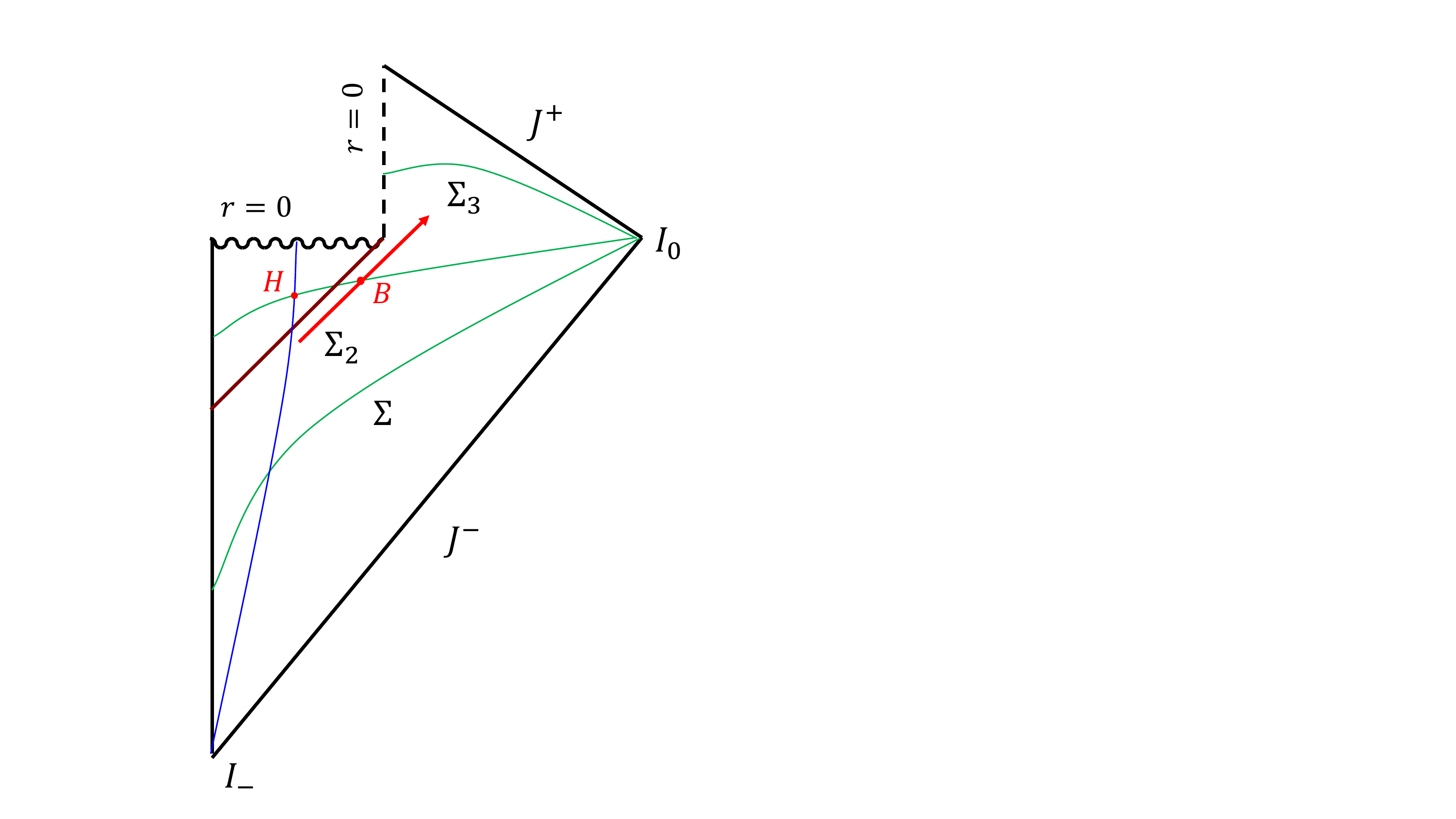}
\caption{Penrose diagram for evaporating black hole.}
\end{figure}

However, there is an assumption here -- namely that these two copies
are independent of each other and that measurements performed at A and
B are independent. And this assumption 
  may be self-contradictory.   For example, if information
evolves in such a way that it is inside, then outside a black hole (or
at $A$ and then $B$), then the points $A$ and $B$ are causally
connected and it is not the case that measurements made at $A$ commute
with those made at $B$. For example, if one imagines that the state
travelled superluminally from $A$ to $B$ and a measurement is made at
$A$ and the same measurement is made at $B$, then from the point of
view of the particle, one expects both measurements to give the same
outcomes. Measurement results at $A$ and $B$ will be correlated, one
should therefore not describe the situation at $A$ and $B$ as
$\ket\psi_A\otimes\ket\psi_B$ (we here work in the low energy limit
where we could potentially describe the Hilbert space as being in a
tensor product).  

Likewise, if Alice performs a measurement inside a black hole on the
in-falling matter, and Bob performs a measurement on outgoing
radiation which encodes that state, then one might expect their
outcomes to be correlated. This is certainly the case if the state has
travelled superluminally from inside of the black hole to outside of the
black hole, and if they perform the same measurement they should
obtain the same result.  Thus rather, than imposing a tensor product
structure, we should say that $B$ is to the future of $A$ and
$\ket\psi_A$ and $\ket\psi_B$ are causally connected or temporally
ordered.  We will argue that this also turns out to be the case for
one effort to solve the black hole information problem, the
Horowitz-Maldacena final state proposal. In this situation, there have
been conflicting claims -- both that the proposal leads to
cloning \cite{PreskillLloyd2014}, and that it does
not \cite{Bennett_slides}.  Here, we will see that since there is a
final state imposed, one can use the Aharonov-Bergman-Lebowitz (ABL)
formula, which is used to calculate the outcome of measurement results
in the case where we have both initial and final conditions. Applying
the ABL formula we find that there are no violations of the no-cloning
theorem.

Another proposed solution to the information problem is black hole complementarity \cite{tHooft-bhcompl,tHooft-bhcompl-string,susskind-bhcompl}. There, one considers two observers, one who is outside the black hole, and one who falls in along with the infalling matter. From the point of view of the external observer, the state of any system remains outside the black hole. If matter is thrown into it, time-dilation causes it to stick to the region just outside the black hole (the stretched horizon) before being evaporated back to the observer. From the external observer's point of view,
the state of the system never crosses the horizon. On the other hand, for the infalling observer, the state does cross the horizon. From a global point of view, we appear to have cloning -- there are
two copies of the state, one inside and one outside. Nonetheless, it was argued that these two observers could never meet up, and thus could never verify that cloning had taken place. One should therefore view these two descriptions as being complementary descriptions. However, black hole complementarity is not sufficient to protect against a violation of entanglement monogamy, which is the central thrust of the AMPS experiment \cite{almheiri2013black}.

Here we will argue that in some situations,
one has a choice whether to impose a tensor product structure on $A$ and $B$, or instead
imagine that A and B are causally connected (we will say that we impose a {\it temporal product} structure between $A$ and $B$).
In the latter case, the apparent violations of cloning and monogamy which appear in the context of the black hole information problem
are of the sort which we indeed can see in quantum mechanics, and therefore need not lead to any contradictions of quantum
theory itself.
Although it may appear otherwise, imposing the temporal product rather than the tensor product, actually does less violence to the theory, since it results in strictly weaker consequences.
This is because imposing a tensor product structure between $A$ and $B$ is a far more radical proposal than imposing a temporal product structure, because it implies both cloning (in the case of the original black hole information problem), and a violation of the monogamy of entanglement (in the case of the AMPS experiment). And both cloning
and entanglement non-monogamy imply super-luminal signaling \cite{toner2006monogamy,oppenheim2014firewalls}. Thus, if one was
aiming to preserve the causal structure of the space-time by imposing a tensor product structure between the inside and outside of the black hole, then one won't succeed, and one will anyway have a break down of the causal structure. 

The issues are sharper in the AMPS experiment, 
which we now summarise. Recall that
to violate the no-cloning
theorem, we need to clone an unknown state. One way to prepare an unknown state, is to prepare a maximally entangled state on two systems $B$ and $R$, and use one of the systems $B$ as an input into the cloning machine which should produce the same state on $A$ while preserving the state
on $B$. The state on $B$ is maximally mixed (unknown), yet if the machine could clone, one would still be left with the maximally entangled state on $BR$ but also $AR$ would be maximally entangled as well -- cloning implies a violation of the principle of entanglement monogamy. Now imagine
we have a black hole which is maximally entangled with some system $R$, either because the black hole started off in a pure state
and then evaporated to half its size, in which case it is now entangled with its emitted radiation $R$ (we say that the black hole has evaporated past its ''Page time''). Or because (to avoid
the issue of computational complexity raised in \cite{harlow2013quantum}), it has been created that way from the start \cite{oppenheim2014firewalls}. Then if the black hole is unitary, it must evaporate in such a way that at the end, the system $R$ and whatever is emitted from the black hole is in a pure state. This means that when the black hole is maximally entangled with $R$,
each Hawking photon $B$ which is emitted should be maximally entangled with $R$. However, this contradicts the fact that each Hawking photon, is also maximally entangled with its infalling partner $A$ -- a violation of the principle of monogamy of entanglement. An observer who falls through the event horizon, carrying the part of $R$ which is entangled with some outgoing radiation $B$, can witness this violation, by performing measurements on $BR$ and $A$ (or just by witnessing that the horizon near $AB$ is unremarkable and satisfies low energy quantum field theory) . One can attempt to invoke black hole complementarity, in the hopes that no observer can witness the violation of entanglement monogamy on $ABR$. However, because the
violation happens for every photon emitted after the Page time, an observer who attempts to jump into the black hole to witness the violation, does not need to collect large amounts of radiation beforehand -- she only needs to examine any Hawking photon. If one wishes to preserve unitarity, it appears one will witness a violation of the principle of entanglement monogamy.

In the original AMPS thought experiment, it was argued that the violation of monogamy of entanglement is such that $B$ is entangled with both $A$ and $R$ (we say that $B$ is polygamous since
one system is entangled with many).
We will find that other situations are possible, where additionally, $AR$ is entangled. We say that $ABR$ is in a sense, polyamorous, in that all three systems are correlated in a more complicated
way than just one system entangled with many. We further find that such a situation is in fact tamer, in the sense that measurement results on systems $A$, $B$ $R$
do not result in a break-down of causality as they do in the original AMPS, and are in fact consistent with quantum theory. This is discsused in Subsection \ref{ss:ampsmeas}.



Let us emphasize   again   however, that in this paper we do not aim to solve the black hole information problem.
Rather, 
if we insist on unitary evolution, what we appear to be sacrificing is the causal structure of general relativity. We need not sacrifice strictly stronger violations of quantum theory such as the no-cloning theorem and monogamy of entanglement. Of course, if we believe information is 
destroyed~\cite{bps,unruh-wald-onbps,OR-intrinsic,unruh2012decoherence}, then no such problems exist. 

We will make use of the fact that measurements on each subsystem of an entangled EPR state such as
\begin{align}
\ket{\Phi^+}_{AB}=\left(\ket{00}_{AB}+\ket{11}_{AB}\right)/\sqrt{2}
\end{align}
give the same statistics as if a system in the maximally mixed state $\id/2$ is first measured at location $A$, and then measured at $B$. This holds up to a unitary on either side for any other maximally entangled state. Thus any two measurement made sequentially is equivalent to two measurements made on an entangled state, even though this
does not hold for a tomographically complete set of measurements\cite{ried2015quantum}.
One could denote such a scenario as  $\id_A/2\temp\id_B/2$
(the temporal product). Thus if the system at A and B is maximally mixed, one gets the same measurement statistics on entangled states where the tensor product is assumed, as on temporally ordered states. In the former case, the measurement statistics can 
violate a Bell inequality, while in the latter case, the statistics violate the Leggett-Garg inequality \cite{LeggettGarg1985}. However, the correlations are the same, it's just that the interpretation is different. In the case of the AMPS experiment, one has maximal entanglement between late time 
outgoing and in-falling radiation, which violates monogamy of entanglement because the late time
outgoing radiation also needs to be entangled with the early time outgoing radiation.
However, if instead we view the outgoing radiation as encoding information which was inside the black hole, and hence, temporally ordered with information inside, then we see that the correlations are allowed by
quantum theory.

To make the discussion of the AMPS experiment and possible violations of monogamy concrete, we will discuss it in the context of the  Horowitz Maldacena (HM) proposal, where it is postulated that 
at the singularity, the state is post-selected in an maximally entangled state. We can imagine that a measurement in a maximally entangled basis happens, with only one outcome being possible. 
This results in post-selected teleportation, transmitting the state inside the black hole, to the outside of the
black hole without the need for any correction to be applied by the person outside the black hole.
Preskill and Lloyd \cite{PreskillLloyd2014} suggested that in the HM proposal, there is violation of monogamy, and violation of the no-cloning theorem. But how should we interpret two measurements, one made inside and one outside 
the black hole, on the state which is teleported through the causal horizon?

Here, we will see that one need not treat those two measurements as commuting, because the teleportation means
that the two measurements are not causally disconnected from each other. 
In this paper, we reinterpret the HM proposal by using the Aharonov-Bergman-Lebowitz (ABL) approach \cite{Aharonov1964}. 
to show that there need not be any violation of the no-cloning principle, nor violation
of the monogamy of entanglement. The ABL formula can be used to calculate the probabilities of measurement outcomes when one has 
both an initial and final condition. As such, it can be applied the the HM proposal, and we will show that the 
proposal is equivalent to  super-luminal particles which can thus travel across the black hole horizon. 
 Our use of a super-luminal particle is intended merely for illustration,  
since it allows us to easily calculate the effect of postselection. Indeed, this gives a natural explanation of why in the HM model 
a ``violation of chronology'' can occur, as analysed in \cite{PreskillLloyd2014}.
One can however speculate that this breakdown of causal structure is a possible alternative proposal to the problem of apparent information loss.

   More generally, we believe that self-consistency is possible, within models that make promises about future events (i.e., final boundary conditions) -- such as the HM model of post-selection at a singularity -- provided that all such promises are constrained to eventually ``fix'' apparent violations of causal order.  

We will see that the HM model is actually isomorphic to a temporal product picture. The isomorphism is given by a mapping 
provided in \cite{PreskillLloyd2014} (called  by the authors ``straightened evolution''). In that paper, the isomorphism  does not have any physical meaning, it is just a mathematical tool.
Our first observation is that this mathematical isomorphism, can acquire operational meaning, if we assume that 
the in-falling radiation in the HM picture is super-luminal. Under such an assumption, it turns out that applying the isomorphism is nothing but 
passing to a new  reference frame such that three particles seen in one reference frame (the HM picture) 
become a single particle seen at different instants of time in another reference frame.
Now, even though in the original HM proposal, none of the  particles are super-luminal, still, {\it all  experiments} 
regarding at different spatial locations will produce statistics that can be mapped 1-1 onto statistics 
coming from experiments performed on a single particle which is super-luminal at some point. 

We could have actually performed our analysis concerning cloning, violation of monogamy  without referring to 
this physical interpretation of the isomorphism, and therefore without introducing super-luminal particles. 
However, we will use the picture, because it is much easier to pass between two pictures, if we have in mind the physical scenario of changing 
reference frames. It is also much easier to understand the problems with loss of ``chronological order'' 
found in Horowitz-Maldacena proposal, as one immediately associates it with the Tolman paradox for superluminal 
particles. The picture has additional attractive, as well as unattractive features, which we will discuss in the Conclusion.

We should emphasize here, that   while   
Lloyd and Preskill  \cite{PreskillLloyd2014}  write that the HM proposal  admits cloning, 
they also state explicitly that , in ``straightened evolution,''  one clone is a future of the other one. 
In our paper we will argue that even in HM proposal itself, cloning does not occur.



Our paper is organized as follows: 
First in Section \ref{sec:super} we show how possible problems with cloning and monogamy map to the temporal product picture (with three particles being 
actually one particle in different instants of time) where it is clearly seen, 
that there is no threat to quantum mechanics. Then in Section \ref{sec:cloningpoly} we examine cloning and possible violations of entanglement monogamy, in the
original the three-particle HM picture, and show that 
due to post-selection, the problems of cloning and violation of monogamy also are of the kind which are in agreement with quantum mechanics.
In Subsection \ref{ss:ampsmeas} we apply this to the AMPS experiment, and show that the original entanglement structure gets modified in the temporal product picture
in such a way that measurements on individual systems do not lead to a breakdown of causality as they do in the original AMPS picture.

Second, in Section \ref{sec:unitarity} we analyse in more depth, the question of  Lloyd and Preskill, of possible interactions between the particles in the
HM proposal. Here, the picture of changing reference frames has an attractive feature, since one can impose that, from the point of view
of the frame where there is a single particle, the evolution should be unitary (or, that ``straightened evolution'' should be unitary). 
We discuss what properties of interactions satisfy this postulate. In Section \ref{sec:complementarity} we see that in the temporal product proposal,
black hole complementarity becomes a mechanism to protect causality in the AMPS experiment. We conclude in Section \ref{sec:conclusion}.

\section{ temporal product $=$  Post-selection $+$ entanglement}
\label{sec:super}

\begin{figure}
\scalebox{0.6}{\includegraphics{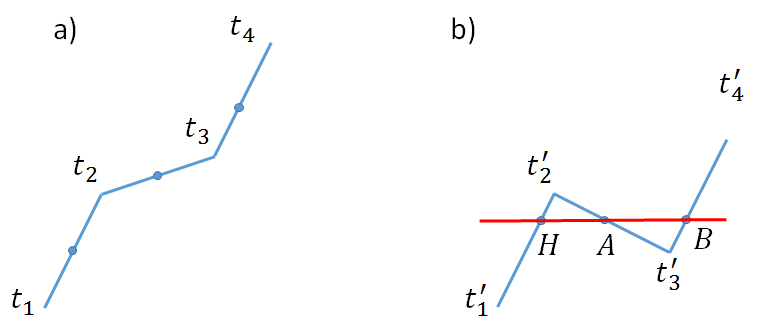}}
\caption{a) in this reference frame, the observer sees one particle, and $H$,$\AA$,${\BB}$ refer to the same system in different instances of time
b) in this reference frame, there are three particles at some time slices. For example, the three particles can be two entangled particles created at $t_3'$ and travelling along paths $\AA$ and ${\BB}$, and one at $H$. A final post-selection occurs at $t_2'$.  }
\label{fig:one_three}
\end{figure}

\begin{figure}
\scalebox{0.6}{\includegraphics{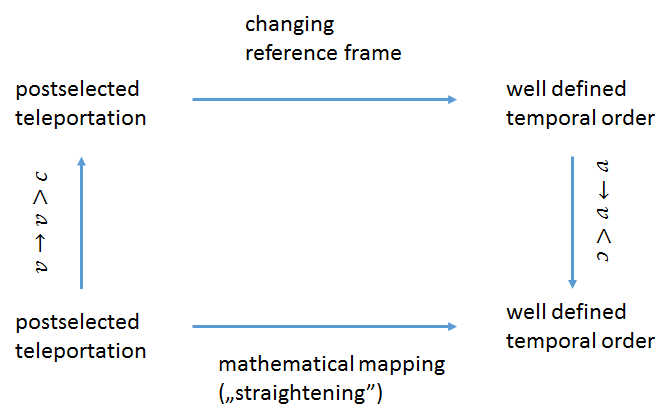}}
\caption{Physical interpretation of Preskill-Lloyd mapping as change of reference frame in the case of super-luminal particle $\AA$.}
\label{fig:diagram}
\end{figure}

In this section we will see that if we have a particle at two space-like separated  points which are related by a temporal product -- i.e. we can obtain the state of the particle at one point, just by unitarily evolving the state from the other space-time point (the particle's past) -- then
this is equivalent to having entanglement between these space-time points and performing a post-selection. The equivalence follows from
a change of reference frame, mapping a super-luminal particle to the HM picture. 
This allows us to obtain the mapping of Lloyd and Preskill \cite{PreskillLloyd2014}, yet we will obtain it not in a formal way,
but in a physical way. 
This is illustrated in Fig. \ref{fig:diagram}. 

A clear example of a temporal product structure, is a particle which travels superluminally.  Let us show that the superluminal particle can
be viewed as postselected teleportation in another reference frame. We describe a particle which  travels for  
some time super-luminally from the perspective of two observers -- one for which the particle moves always into the future and one for which the particle moves for some time into the past.
Let $t_1 < t_2 < t_3 < t_4$ be ordered times.
Consider a particle traveling to the right, which has sub-luminal velocity in period $(t_1,t_2)$, then super-luminal in period $(t_2,t_3)$ and finally again sub-luminal in period $(t_3,t_4)$. There exists a reference frame, where the order is $t'_{1}<t'_{3}<t'_{2}<t'_{4}$, 
i.e. the order of $t_2$ and $t_3$ is reversed.  In such a reference frame instead of one particle, 
we have three particles, two traveling to the right, and one (antiparticle) traveling to the left, see Fig. \ref{fig:one_three}.

Evolution of the particle in the original reference frame can be written as 
\be
|\Psi(t_4)\rangle_H=U(t_4,t_1) |\psi\>_H=U(t_4,t_3)_HU(t_3,t_2)_HU(t_2,t_1)_H|\Psi(t_1)\rangle_H.
\ee
Let us now find the natural description of evolution in a reference frame where there are three particles.
We start with equation
\be
|\Psi(t'_4)\rangle_H=U(t'_4,t'_1)_H |\psi\>_H=U(t'_4,t'_3)_HU(t'_3,t'_2)_HU(t'_2,t'_1)_H|\Psi(t'_1)\rangle_H
\ee
Next we put between any two unitary operators resolution of the identity $I_H=\sum_i|i \rangle\langle i|_H$
\begin{eqnarray}
&& |\Psi(t'_4)\rangle_H = U(t'_4,t'_3)_HU(t'_3,t'_2)_HU(t'_2,t'_1)_H|\Psi(t'_1)\rangle_H=\nonumber\\
&& =\sum_{i,j,k,l}|i\rangle\langle i|_HU(t'_4,t'_3)_H|j\rangle\langle j|_HU(t'_3,t'_2)_H|k\rangle\langle k|_HU(t'_2,t'_1)|l\rangle\langle l|\Psi(t'_1)\rangle_H.
\end{eqnarray}
Finally we use the mathematical identity
\begin{eqnarray}
&& \langle i|_HU(t'_4,t'_3)_H|j\rangle_H\langle j|_HU(t'_3,t'_2)_H|k\rangle_H\langle k|_HU(t'_2,t'_1)_H|l\rangle_H=\nonumber\\
&& =\langle i|_{\BB} \otimes \langle j|_{\AA} \otimes \langle k|_H U(t'_4,t'_3)_{\BB} \otimes U(t'_3,t'_2)_{\AA} \otimes U(t'_2,t'_1)_H|j\rangle_{\BB} \otimes |k\rangle_{\AA} \otimes |l\rangle_H=\nonumber\\
&& =\langle i|_{\BB}\otimes \langle k|_{\AA}\otimes \langle k|_H U(t'_4,t'_3)_{\BB} \otimes U(t'_3,t'_2)^{T}_{\AA} \otimes U(t'_2,t'_1)_H|j\rangle_{\BB} \otimes |j\rangle_{\AA} \otimes |l\rangle_H.
\end{eqnarray}
and rewrite Eq. (2) in the  form
\begin{eqnarray}
&& |\Psi(t'_4)\rangle_{\BB}=\nonumber\\
&& =\langle \Phi^+|_{{\AA}H}U(t'_4,t'_3)_{{\BB}} \otimes U(t'_3,t'_2)^{T}_{\AA} \otimes U(t'_2,t'_1)_H|\Phi^+\rangle_{{\BB}{\AA}} \otimes |\Psi(t'_1)\rangle_H,
\end{eqnarray}
where
\begin{eqnarray}
|\Phi^+\rangle_{XY}=\sum_i|i\rangle_X |i\rangle_Y
\end{eqnarray}
is a (supernormalized) maximally entangled state.

We interpret it in the following way. At time $t_3'$ a pair of particles $\AA$ and ${\BB}$ in the maximally entangled state is created. 
Then at time $t_2'$ particles $H$ and $\AA$ are projected onto the maximally entangled state and annihilate.
We also note that in place of unitary operators we can put arbitrary operators $X$, $Y$, $Z$ and obtain
\begin{eqnarray}
X_HY_HZ_H|\Psi(t'_1)\rangle_H\rangle=\langle \Phi^+|_{{\AA}H}Z_{{\BB}} \otimes Y^{T}_{\AA} \otimes X_H|\Phi^+\rangle_{{\BB}{\AA}} \otimes |\Psi(t'_1)\rangle_H
\end{eqnarray}
where we now have via a change of reference frame, gone from a single particle, to three particles and post-selected teleportation.

The picture where we have a single particle clearly corresponds to  {\it a temporal product}, since the particle at $t_4$ is just the particle at $t_1$, evolved in time. One might have thought
that the picture where we have three particles should be a tensor product picture. However because it is equivalent to the single-particle picture, due to post-selected teleportation,
the three-particle picture is isomorphhic to the single particle one, and therefore effectively also corresponds to a temporal product. We will explore this connection more fully,     and   it
will become clear that the HM proposal is a cryptic form of temporal product, 
  allowing it to avoid any   
 ``firewall versus monogamy   confrontation.  

Note that in such a pre and post-selected
scenario, the ABL formula gives a simple way to calculate the probability of measurements made in between the times $t_i$ and $t_f$ of a pre- and post-selection. Since we can treat preparations
as measurements, we imagine that at $t_i$, there is an
initial measurement/preparation represented by the projector $\Pi_i$ (in this case, the projector onto $| \Phi^+\rangle_{{\BB}{\AA}}$), and the post-selection is represented by a measurement
whose outcome is given by the successful projection onto the projector $\Pi_f$ (in this case, the projector onto $| \Phi^+\rangle_{H{\AA}}$). 
  We may then   calculate the probability of measurement results made
in between $t_i$ and $t_f$, represented by a set of projectors $\{P_k\}$.   In particular,   the probability of obtaining the sequence of measurement outcomes corresponding to $\Pi_i$ followed by $P_k$, followed by $\Pi_f$ on the state $\rho$ is given by $\tr \Pi_f P_k \Pi_i\rho$, and thus the probability $p(k|\Pi_i,\Pi_f)$, of obtaining the intermediate measurement outcome corresponding to $P_k$ given the pre and post selection corresponding to $\Pi_i$ and $\Pi_f$, is
\begin{align}
  \label{eq:abl}
    p(k|\Pi_i,\Pi_f) =  \frac{\tr (\Pi_f P_k \Pi_i\rho)}{\sum_k\tr (\Pi_f P_i \Pi_i\rho)}
\end{align}
i.e. just the probability of the sequence of measurements corresponding to intermediate outcome
$k$ normalised by the probability of obtaining any intermediate outcome.  This is the ABL formula.

\section{Cloning and  non-monogamy in a post-selected world}
\label{sec:cloningpoly}
\subsection{In a black hole, the clones are not independent (and so are not clones)}
\label{ss:cloning}

Let us see now that in both the picture where we have post-selection, and equivalently, in the picture where we have a superluminal particle, the cloning is only apparent. In other words, let us see
that measurement outcomes made at $H$ and at ${\BB}$ will be correlated, which should clearly not be the case if they were independent clones. In fact, the particle at ${\BB}$ is just the future version of the particle at $H$. While this violates the causal structure of the space time, it is a strictly weaker implication than cloning, which also allows one to signal superluminally and violate the causal structure.
In the superluminal particle picture case our approach  is compatible with Sec. 3.1 of Lloyd and Preskill, 
where they state, that one clone is future of the other clone,  but  our description of cloning in the HM proposal goes beyond 
their analysis.

\begin{figure}
\scalebox{0.4}{\includegraphics{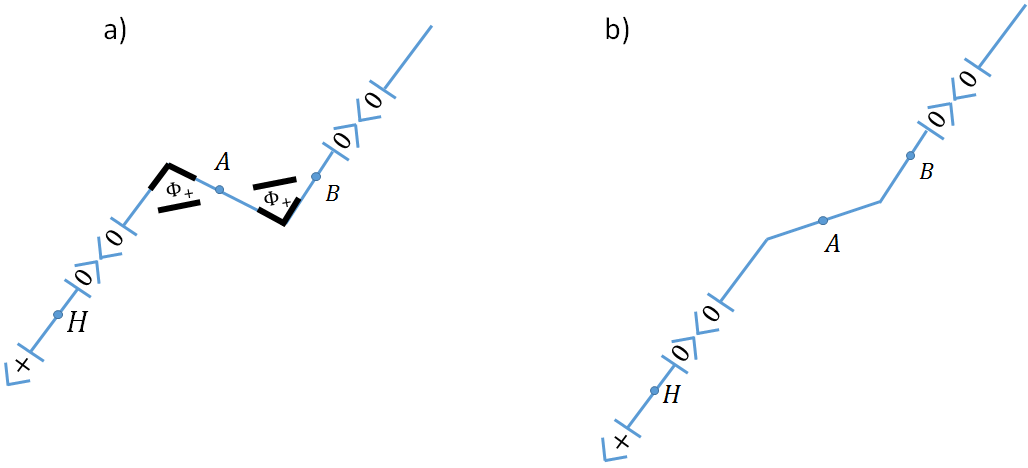}}
\caption{Measurement testing independence of clones viewed in two different reference frames (as in Fig. \ref{fig:one_three} $H$,${\AA}$,${\BB}$ are seen by one observer as three systems, while by another observer as one system
appearing at different instance of time.) \label{fig:cloning}
}
\end{figure}

We can see that the clones are correlated   in the picture with post-selection   (see Fig. \ref{fig:cloning}),   by using  
the ABL formula Equation \eqref{eq:abl} to compute the probability of measurement outcomes. 
Let us suppose that at time slice $t_0$ we have three particles -- the particle $H$ is in a state $|\Psi\rangle$ while the particles $\AA$ and ${\BB}$ are in the maximally entangled state $|\Phi^+\rangle$. At a later time $t_1$ the particles $H$ and $\AA$ are always projected on the maximally entangled state $|\Phi^+\rangle$ -- this projection serves as a final boundary condition. Hence we conclude that at time $t_0$ the particle ${\BB}$ is also in a state $|\Psi\rangle$.
It looks like we have cloning. Indeed at some time slice, two particles in the same state appear
at two different places. 

It is instructive to see what this looks like in the reference frame where there is only one particle, i.e. if we transform from the tensor product picture to the temporal product picture (see Fig. \ref{fig:cloning}). 
There, the three particles are simply subsequent time instances of the same particle. 
Why  do we not regard them as clones? Because clones have to be independent, whereas the states of a single particle 
in subsequent instances of time are not.   For example,   suppose the initial state of particle $H$ is $|+\>$.
If we  measure the particle at $H$ in the computational basis, and  project it onto $|0\>$,
then the state of the particle at later times will collapse to $|0\>$. 
Thus, at a subsequent time, the probability of obtaining $|0\>$ is one, while the probability of obtaining $|1\>$ 
is zero. The probability of obtaining $|0\>,|0\>$ (or $|1\>,|1\>$ at both times is therefore $1/2$. Hence, we conclude that the ''clones'' are not independent copies, but are in fact highly correlated with
each other. This was actually noted by Bennett \cite{Bennett_slides}, that in post-selected teleportation we do not have cloning.

Let us see how this translates into the three particle picture with post-selected teleportation. Some measurement outcomes are impossible.
Consider for example, two particles $H$ and $\AA$, and suppose, 
that by measuring $H$, one obtained $|0\>$ and by measuring ${\BB}$ one obtained $|1\>$.
Due to the latter measurement, the particle $\AA$ will collapse to the state $|1\>$, 
and therefore  the probability of obtaining the post-selection $|\Phi^+\>_{HB}$
will be zero because the latter has even parity. Thus such an event cannot occur in post-selected scenario, because in such scenario,
all statistics must be compatible with the final boundary condition. 
Thus the probability of obtaining $|0\>$  on particle $H$ and $|1\>$ on $\AA$ is zero, precisely as in the case of the single superluminal particle and unlike the case of two clones.
This must be   so,   because the two pictures are equivalent. So we have learned that 
in the post-selected scenario one has to be careful: since we impose not only initial but also final conditions, not all measurement results can occur.   In this way the ``promise'' of the final condition imposes consistency between the outcomes of intermediate measurements, such that apparent `cloning' correlations are no more problematic than standard temporal correlations.   

 We can use the the ABL formula \eqref{eq:abl} to   check   that the probabilities of measurement outcomes $M_1$ and $M_2$ are the same as in the single superluminal particle case (and completely correlated). For illustrative purposes, let us now explicitly calculate the  probability of obtaining $|0\rangle$ in 
both measurements, conditioned on obtaining  $|\Phi^+\>$ as a final state of particles $H$ and $\AA$ (see Fig. \ref{fig:circuit}):
\begin{figure}
\scalebox{0.8}{\includegraphics{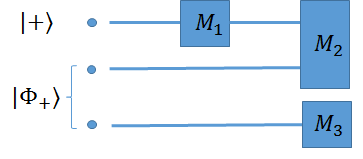}}
\caption{Measurement scheme to probe ``clones''.
  The measurement $M_1$ has projectors $P_0=|0\>\<0|,P_1=|1\>\<1|$, and prepares $|0\>$  or $|1\>$ respectively; The
  with outcomes $0,1$
  $M_2$ is a measurement in the Bell basis, and we are interested in the post-selection corresponding to the projector $\Pi_{\Phi_+}=|\Phi_+\>\<\Phi_+|$ which we denote by $M_2=\Phi_+$.
$M_3$ is also in the basis $|0\>,|1\>$}
\label{fig:circuit}
\end{figure}
The joint probability of obtaining $|0\>$ in both measurements  and  final projection
$\Pi_{\Phi_+}$ given that the initial state is $ |+\> \otimes |\Phi^+\>$ is
\ben
p(M_2=\Phi^+, M_1=0,M_3=0)=\tr (P_0\otimes\id \Pi_{\Phi+}\otimes P_0|+\>\<+| \otimes |\Phi^+\>\<\Phi^+|)
=\frac18
\een
The probability of obtaining $|\Phi_+\rangle$ in  measurement $M_2$ regardless of the outcome of the measurement $M_2$ is
\be
p(M_2=\Phi^+)=p(\Phi^+,0,0)+p(\Phi^+,0,1)+p(\Phi^+,1,0)+p(\Phi^+,1,1)=\frac18+0+0+\frac18=\frac14.
\ee
where $p(\Phi^+,i,j)= p(M_1 =i, M_2=\Phi^+,M_3=j)$ 
Thus the probability of obtaining $|0\>$ in
both measurements with post-selection onto $|\Phi^+\>$ is 
\be
p(M_1=0,M_3=0|M_2=\Phi^+)= \frac{p(M_1=0,M_2=\Phi^+,M_3=0)}{p(M_2=\Phi^+)}=\frac{\frac18}{\frac14}=\frac12
\ee
which is the same as we obtained in the one-particle picture, as it should be. 
  This   shows that the clones are not independent. If we had two true clones, the state would be $|+\>|+\>$,
and the probability of obtaining the $|0\rangle$ outcome for both clones would be $1/4$. 

\subsection{Polyamory in space versus polyamory in time}
\label{ss:intime}

We have just seen that what appears to be cloning need not be, because the clones are correlated
and so are not independent copies on a tensor product space: rather, one is just the temporally evolved state of the other. We will now see that what appears to be a violation of monogamy of entanglement is actually just polyamory in time, which is allowed by quantum theory and so again is a strictly weaker implication   of the model.  

Let us suppose that at time slice $t_0$ we have three particles -- particle $H$ is in a state $|\Psi\rangle$ while particles $\AA$ and ${\BB}$ are in the maximally entangled state $|\Phi^+\rangle$. At later time $t_1$ the particles $H$ and $\AA$ are always post-selected onto the maximally entangled state $|\Phi^+\rangle$. Hence, we conclude that at time $t_0$ the particle $\AA$ is maximally entangled both with the particle $H$ and with the particle ${\BB}$ and monogamy of entanglement is violated. However, it is the kind of violation that is allowed  in quantum mechanics.

\begin{figure}
\scalebox{0.6}{\includegraphics{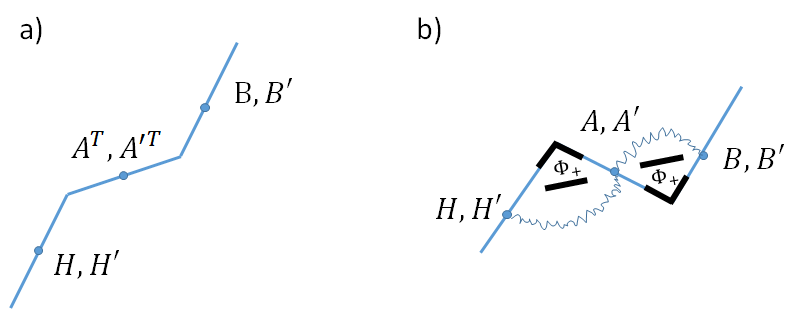}}
\caption{Polyamorous violation of the CHSH inequality. Here, the violation is not of a spatial CHSH inequality, but a temporal CHSH inequality}
\label{fig:Bell}
\end{figure}

Again, it is instructive to look in the one particle picture 
(see Fig. \ref{fig:Bell}). Let us suppose that we prepare a particle in the maximally mixed state and we first measure observable $X$, then 
$Y$, and finally $Z$.
For consecutive measurements of observable $X$ and observable $Y$ on a single particle in the maximally mixed state, the joint probability of outcomes 
$p(x,y|X,Y)$ is the same as the probability of outcomes $p(x,y|X,Y^T)$ for measurement on the maximally entangled state.
This is because measuring $X$ on half of a maximally entangled pairs prepares eigenstates of $X^T$  on system $\AA$. 
The same is true of $p(y,z|YZ)$ [but not of $p(a,c|XZ)$], as the measurement of $Y$ disturbs the system, 
and measurement of $X$ does not prepare eigenstates of $X^T$ on system ${\BB}$. The correlations
which arise from these measurements are well known, and we say that they violate a 
{\it temporal Bell inequality} \cite{LeggettGarg1985}. Such correlations do not
describe a violation of monogamy (even though the statistics are similar), because the measurements are made sequentially. We thus do not have a violation of monogamy of spatial entanglement, but rather, polyamory in time, which is allowed in quantum theory.

Now we can pass to the picture of three particles where post-selection will   (and must)   give us the same answer as in the single particle picture. 
We prepare the particle $H$ in the maximally 
mixed state. Then due to post-selection, the measurements on $H$ and $\AA$ 
will be correlated as if they were performed on the maximally entangled state. The measurements on $\AA$ and ${\BB}$ 
will be similarly correlated, because particles $\AA$ and ${\BB}$ were prepared in the state $|\Phi_+\>$.
Thus we can violate Bell inequalities between $H$ and $\AA$ and between $\AA$ and ${\BB}$, by choosing Alice's CHSH measurements on $H$ particle, Bob's CHSH measurements on $\AA$, and Charlie's CHSH measurements on the particle at ${\BB}$.

What happens if, instead of the maximally mixed state, we prepare the particle in some other state? Let us again look at the single particle picture. 
Without loss of generality let us suppose that we prepare the particle in a state $|0\>$ and measure the Pauli matrix $\sigma_z$. Then only result $|0\>$ will occur. On the other hand for the maximally mixed state both results $|0\>$ and $|1\>$ will occur. Hence, in general, for consecutive measurements of observable $X$ and observable $Y$ on a single particle the joint probability of outcomes $p(x,y|X,Y)$ is not the same as the probability of outcomes $p(x,y|X,Y^T)$ performed on the maximally entangled state.  However if we choose the CHSH measurements, we still violate a temporal Bell inequality. We also obtain the same statistics for measurements outcomes in three particle picture. 

%
%
%
%
%
%

\subsection{Polyamory vs Polygamy: additional structure of entanglement in the AMPS firewall experiment}
\label{ss:ampsmeas}

\begin{figure}
\scalebox{0.4}{\includegraphics{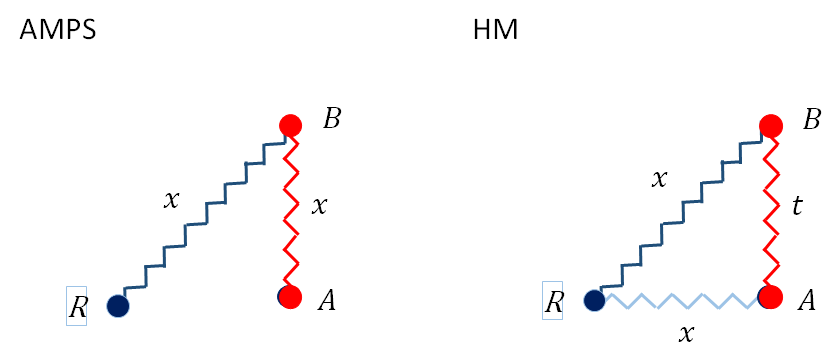}}
\caption{In AMPS (left figure) we have spatial entanglement ({\it x}) between late time outgoing Hawking radiation $B$ and both the early radiation $R$ and its infalling Hawking partner $A$. This violation of entanglement monogamy leads to causality violations using just measurements on the individual systems. If instead the spatial entanglement on $AB$ becomes entanglement in time, then $AR$ also becomes spatially entangled (right figure), and then the correlations obtained do not lead to such paradoxes.  The entanglement structure of the right figure corresponds to a temporal product between $A$ and $B$ as is found in the HM proposal.
}
\label{fig:amps_hm}
\end{figure}

The above considerations allow us to examine and interpret the statistics of measurements made in the AMPS firewall experiment. We will see that in the HM proposal, as well as the single particle picture, there are additional correlations to those considered in the AMPS experiment, namely, temporal correlations between the infalling Hawking radiation and the reference system $R$
(see Figure \ref{fig:amps_hm}). These lead to very different
measurement outcomes when performing the AMPS experiment, and we will see that the violation of entanglement monogamy is much tamer.
We can here consider each system to be a qubit for simplicity.
Recall that AMPS consider the situation where a reference system $R$ is outside the black hole and is maximally entangled with an outgoing Hawking particle $B$.
Then, if there is {\it no drama}
near the black hole horizon (i.e. low energy quantum field theory describes the region near the horizon and there is no firewall),
$B$ must also be maximally entangled with an infalling partner $A$. The entanglement is {\it polygamous} because $B$ is entangled with more than one system.
An observer who falls into the
black hole will witness this violation of the principle of entanglement monogamy by performing measurements on systems $ARB$ to reveal entanglement between $BR$ and $AB$.

In terms of what measurement to perform on $ARB$ to witness a violation, a set of bipartite measurements are discussed in \cite{PreskillLloyd2014}. Here,
we follow \cite{oppenheim2014firewalls} and make measurements which would allow superluminal signalling bewteen $B$ and $AR$.
Let us first consider the standard AMPS case, where one has spatial entanglement between $AB$ and $BR$, and for the purpose of illustration let us consider each to be in maximally entangled state
 $(|00\>+|11\>)/\sqrt{2}$.  Then one can use this to  
signal faster than light between $B$ and $AR$ as follows: If we want to use $B$ to signal a $0$, we measure it in the computational basis, and if we want to communicate $1$, we
measures in the complementarity basis $|\pm\>=(|0\>\pm|1\>)/\sqrt{2}$. Then, because $B$ is maximally entangled with both $A$ and $R$, both $A$ and $R$ will collapse to the same state.
If we measured $B$ in the computational basis, $AR$ will be left in the state
$(|00\>\<00|+|11\>\<11|)/2$ while if we measured in the $|\pm\>$ basis, it will collapse to $(|--\>\<--| +|++\>\<++|)/2$. These two density matrices are not the same, and can be distinguished
with probability $3/4$ via individual measurements on $A$ and $R$. This leads to signalling between $B$ and $AR$, either probabilistically, or with arbitrary certainty by repeating this protocol sufficiently many times.

Essentially, we can regard $ABR$ as arising from entanglement on $BA$ and then cloning the state of $A$ onto $R$, and it is known that such a situation leads to superluminal
signalling \cite{dieks1982communication}. Because $A$ and $R$ are clones of each other, the total state $AR$ is distinguishable depending on what basis $B$ was measured in.
The above protocol is also equivalent to simultaneous violations of a Bell inequality between
$B$ and $R$ as well as between $B$ and $A$ and in particular, the simultaneous violation of the CHSH inequality.  This is not allowed in any no-signalling theory \cite{toner2006monogamy}, thus the
above measurements have the advantage that the violation rules out more than just quantum theory. One also need only measure each system individually.

\begin{figure}
\scalebox{0.4}{\includegraphics{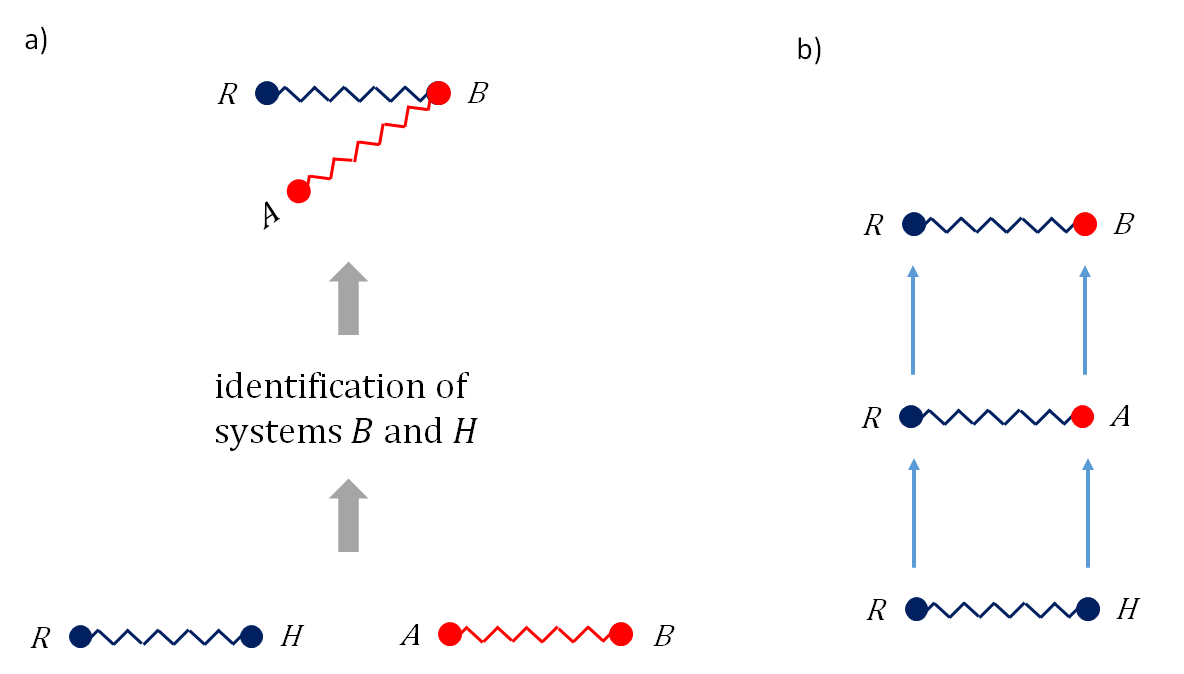}}
\caption{In the AMPS experiment we can label the system inside the black hole which is correlated with a reference system by $H$. When this information leaves the black hole as Hawking radiation, we identify it with the outgoing Hawking radiation $B$. 
Thus in AMPS, the polygamous entanglement comes from identification of system $H$ and $B$, while keeping systems $A$ and $R$ uncorrelated. In the temporal product picture, $H$ and $B$ are identified by $H$ being the past of $A$ which is the past of $B$.}
\label{fig:ampsident}
\end{figure}

On the other hand, if we have a temporal product between $A$ and $B$, then the above breakdown
of causality is resolved. In particular, if the correlations on $AB$ are temporal rather than spatial, then in addition we have spatial entanglement on $AR$ as depicted in
Figure \ref{fig:amps_hm}. These additional correlations prevent signaling from $B$ to $AR$ because in this case
the measurement performed on $A$ influences the state of $BR$ rather than  measurement performed on $B$ influences 
the state of $AR$. Hence we have only superluminal signaling from $A$ to $BR$. 
A comparison of the situation in the tensor product picture vs the temporal product picture is given in Figure
\ref{fig:ampsident}.

\begin{figure}[h]
  \centering
\includegraphics[trim=2cm 6cm 3cm 3cm, clip=true, totalheight=0.2\textheight]{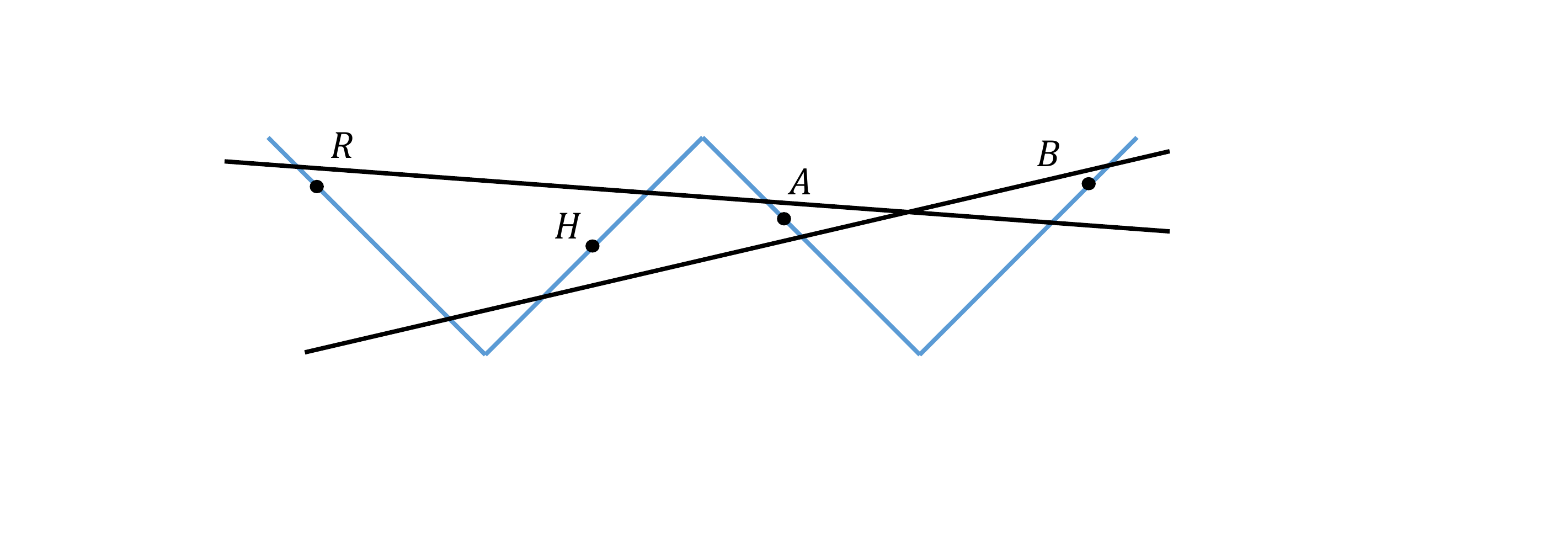}
\caption{We consider the situation considered by AMPS, but in the HM proposal.  System $H$ which is entangled with reference frame $R$, is initially inside the black hole, and is teleported out of it through entanglement formed by the infalling $A$ and outgoing $B$ Hawking pair. For any intermediate measurement made on the systems, the order doesn't matter. }
\label{fig:amp_ps}
\end{figure}

We can calculate the probabilities of measurement outcomes 
made in the HM model, and verify that indeed the entanglement structure depicted
in Figure \ref{fig:amps_hm} is what arises and that there is no violation of quantum mechanics with such a structure. We do this using the ABL rule in the situation depicted in Figure \ref{fig:amp_ps}. In this case, we find that there is no signalling from $B$ to $AR$. This can be seen immediately from
Figure \ref{fig:amp_ps} where it is clear that the order of intermediate measurements on $A$,$B$ and $R$ doesn't matter for calculating probabilities using the ABL rule. A physical explanation comes from Figure \ref{fig:amps_hm}. The singlet on $BR$ is just the future singlet of $AR$. If one imagines the protocol above to signal from $B$ to $AR$, then it doesn't work, because $A$ is in the past of $B$ 
and any measurements made on it will result in breaking the entanglement on $BR$.


\subsection{Teleporting via polyamorous entanglement in time is impossible}
\begin{figure}
\scalebox{0.6}{\includegraphics{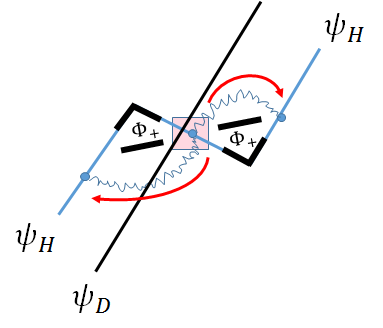}}
\caption{Cloning through polyamorous entanglement?}
\label{fig:tele_clo}
\end{figure}

The polyamorous nature of the entanglement in the three-particle picture 
of Fig. 3 does not provide additional resources for tasks such as teleportation. To see this, consider an attempt to teleport a fourth particle D, as in Fig. 4, where a Bell measurement is made on D and $\AA$. One can expect (wrongly) that there will appear two clones along the entanglement directions.
(see Fig. \ref{fig:tele_clo})

\begin{figure}
\scalebox{0.6}{\includegraphics{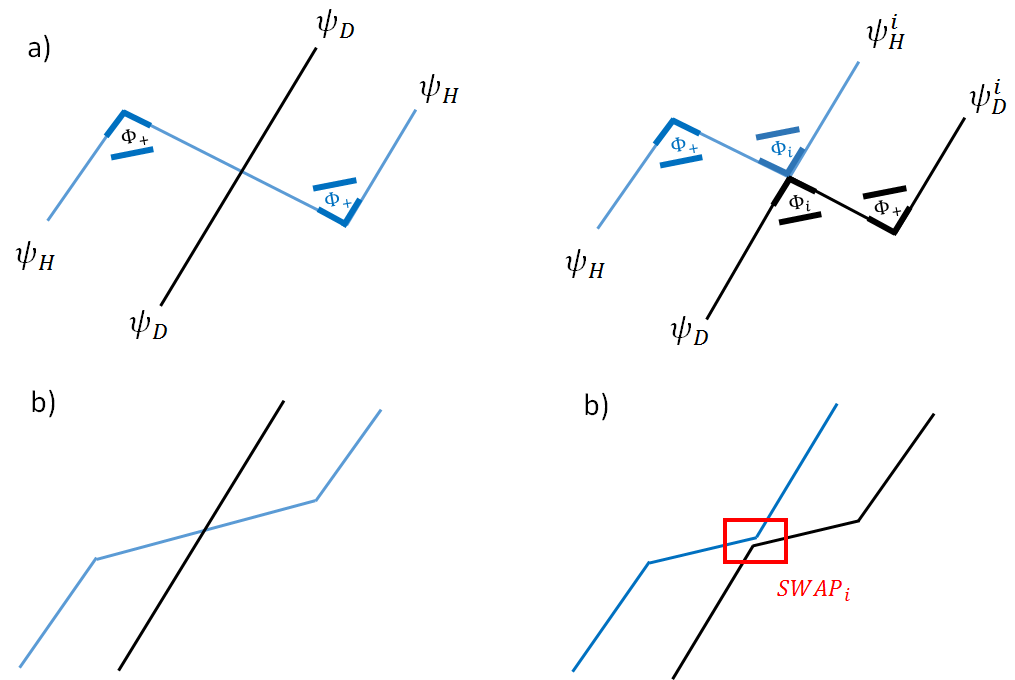}}
\caption{The effect of teleporting a state of a particle through ``polyamorous'' entanglement. a) three particle picture b)  one particle picture}
\label{fig:teleport}
\end{figure}

Let us now see how it looks  in the picture of three particles. 
It is described in 
Fig.~\ref{fig:teleport}a). Namely, the Bell measurement has teleported the particle D, and also created an entangled pair 
which was used in post-selected teleportation of the particle at $H$. As a result we   find   that the states of the particles	$H$ and D 
have been swapped. Thus we do not obtain two clones of $|\psi_D\>$.


The passage from  the three particle 
 to the one particle picture 
 is via a transpose on system $\AA$.
 Thus the Bell measurement projectors become the swap operator composed with a Pauli matrix on the
 $\AA$ system.
As a result, the particles  $D$ and  $H$  simply swap their states.  Again, the state of particle $D$ 
instead of being cloned, is just swapped with the state of particle $H$.

Thus, we have here a violation of monogamy, 
  but of a kind that   is allowed in quantum mechanics  and   which   does not lead to cloning.

  \section{``Unitary'' evolution in a post-selected world}
  \label{sec:unitarity}

We have shown so far, that there is apparently no problem of cloning and polyamory of entanglement in a post-selected world. 
However this does not mean that that we have removed all the problems with the Horowitz-Maldacena proposal. Indeed, Gottesman and Preskill \cite{GottesmanPreskill2004} realised that in a post-selected world, if the particles interact with one another, 
then difficulties may arise. This was developed later by Lloyd and Preskill \cite{PreskillLloyd2014}, who gave further examples of evolutions  that should be prohibited, as they lead to paradoxes such as ''chronology violation'' (which physically 
would correspond to closed time-like curves. That interactions may pose a problem can be seen from the fact that we are attempting to teleport a particle at $H$ through entanglement on ${\AA}{\BB}$, and the final post-selection onto a maximally entangled state on $H{\AA}$ succeeds in teleporting the particle only if $H$ is in a product state with ${\AA}{\BB}$ and if ${\AA}{\BB}$ is in the maximally entangled state. If $H{\AA}$ interact (or if there is any interaction on $H{\AA}{\BB}$), both these conditions will almost certainly be violated. 

This is a significant drawback to the HM proposal, and we know of no satisfactory proposal for fixing it. However, in the picture of a single particle, solutions appear to be less ad hoc. 
Our goal is to find a reasonable evolution in a post-selected world, and we
therefore propose the following postulate.

\noindent
{\bf Evolution postulate:} 
{\it In the reference frame where there is one particle the evolution should be unitary.}

This clearly violates relativity, as it introduces a preferred frame; however, we are already violating the causal structure,  and  this is a strictly weaker violation than cloning and polyamory, as we have pointed out throughout. We will see that it may be a more natural way to avoid the difficulties which plague the HM proposal pointed out by Gottesman-Preskill. Using this postulate   also   lets us characterize the operations which are allowed in a post-selected world.

\subsection{Coupling the system with itself - how to avoid closed time-like curves?}
\label{sec:ctc}

As noted by Gottesman and Preskill and analysed in more detail in \cite{PreskillLloyd2014}, 
significant problems appear if particle $\AA$ interacts with particle ${\BB}$ (i.e. when outgoing radiation interacts with infalling   matter ).  
Namely, one obtains something which can be interpreted as a closed timelike path --  particle ${\BB}$ (future) may influence  particle $\AA$
 (past), which happens, e.g. if the interaction is a C-NOT gate (with ${\BB}$ as source and $\AA$ as target). 


Below we will characterize all interactions between $\AA$ and ${\BB}$ (i.e. interactions across the horizon) 
as well as $H$ and $\AA$ (i.e. infalling matter with infalling radiation),   that are allowed by   our evolution postulate. 
We will then argue that these are simply the ones that preserve the maximally entangled state. We will also characterize   the allowed   interactions 
between particles $H$ and $\AA$ (i.e. infalling matter with outgoing radiation).

{\it Interactions between $\AA$ and ${\BB}$ and  between  $H$ and $\AA$.} In order to characterize what operations are allowed, we assume that particle $\AA$ interacts with particle ${\BB}$ via a (not necessarily unitary) operation $W_{{\AA}{\BB}}$ and require that the final state of particle ${\BB}$ is related to the initial state of particle $H$ via a unitary operation 
$U_{H\rightarrow {\BB}}$. We can compute the effect of the operation and postselection onto $|\Phi^+\>_{H{\AA}}$ as follows:


\begin{eqnarray}
U_{H\rightarrow {\BB}}&=&\<\Phi^+|_{H{\AA}}W_{{\AA}{\BB}}|\Phi^+\>_{{\AA}{\BB}}=\nonumber\\
& =&\<\Phi^+|_{H{\AA}}\sum_{ijkl}w_{ijkl}|i\>_{{\AA}}|j\>_{{\BB}}\<k|_{{\AA}}\<l|_{{\BB}}|\Phi^+\>_{{\AA}{\BB}}=\nonumber\\
& =&\sum_{ijk}w_{ijkk}|j\>_{{\BB}}\<i|_{H}
\end{eqnarray}

This can be written in the form
\begin{eqnarray}
 \tr_{\AA}(V_{{\AA}{\BB}}W_{{\AA}{\BB}}^{T_{\AA}}) = U_{H\rightarrow {\BB}}
\label{unitarylhs}
\end{eqnarray}
where
\begin{eqnarray}
\label{unitarybc}
V_{{\AA}{\BB}}=\sum_{ijkl}|i\>_{{\AA}}|j\>_{{\BB}}\<j|_{{\AA}}\<i|_{{\BB}}
\end{eqnarray}
is the operator which swaps the states of particles $\AA$ and ${\BB}$. We conclude that any operation $W_{H{\AA}}$, such that the left hand side of Equation \eqref{unitarylhs} is a unitary, is allowed.
We also note that similar reasoning can be applied to the interaction of particle $H$ with particle $\AA$.

One can check  that the above condition is equivalent  to the condition:
\be
W_{{\AA}{\BB}}|\Phi^+\> = |\Phi_{\text{max ent}}\>
\ee
where $|\Phi_{\text{max ent}}\>$ is a maximally entangled state i.e. $|\Phi_{\text{max ent}}\>=U\ot I |\Phi^+\rangle$ for some unitary $U$.
One can easily understand why this must be so. Clearly, such $W$ does not disturb teleportation, since the resulting state is still maximally entangled, 
and hence the condition is sufficient.   Conversely,   if $W$ would transform $\Phi^+$ into a state which is not maximally 
entangled, teleportation cannot be faithful anymore, and the state of particle $H$ will not be unitarily related 
to particle ${\BB}$.

By the same argument, any interaction $W_{H{\AA}}$ between particles $H$ and $\AA$ satisfies our   evolution postulate   if and only if 
$W_{H{\AA}}^\dagger |\Phi^+\rangle$  is maximally entangled. Finally, note that   neither of   $W_{H{\AA}}$ and $W_{{\AA}{\BB}}$ need  be unitary.
Thus, in a picture with a super-luminal particle, what looks like a standard unitary evolution for one observer 
(the one who sees a single particle), can be perceived  as non-unitary by another observer.  


\subsection{Coupling with another system: Destruction of information}

\begin{figure}
\scalebox{0.5}{\includegraphics{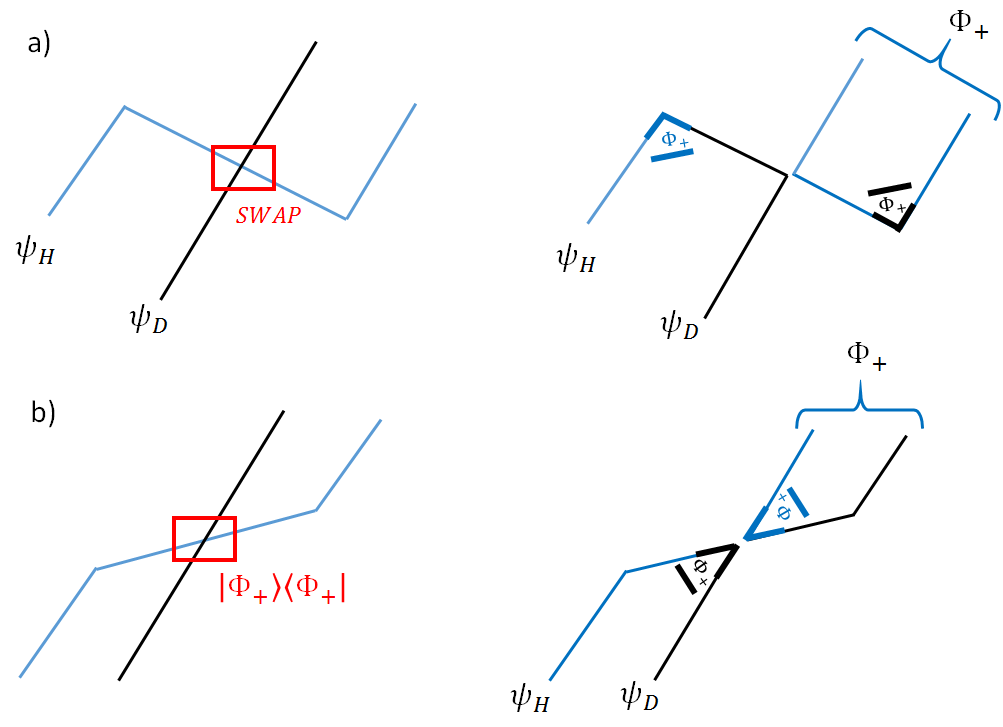}}
\caption{The swap in the post-selected world destroys information: it maps arbitrary states to final state $\phi_+$.}
\label{fig:destr}
\end{figure}

Suppose that in a post-selected world, a fourth particle $D$ interacts with particle $\AA$ just via a the unitary operation swap. Then in the one-particle picture, the unitary is mapped to the maximally entangled projector, which changes arbitrary input states 
$\psi_H$ and $\psi_D$ into joint maximally entangled state (see Fig. \ref{fig:destr}a) and thus leads to information destruction. The same can be 
seen in the three particle picture (Fig. \ref{fig:destr}b).

The solution proposed by Preskill and Lloyd is the following: in a post-selected world,
only those unitary operations acting on 
  particles B and D in   Figure \ref{fig:destr} are allowed, which after partial transpose are again unitary. 
Here we will see that our evolution postulate goes a bit further:  even a {\it non-unitary} operation is fine, 
provided its partial transpose is unitary.

Let us now characterize,   more generally,   all operations which act on particle $\AA$ and particle $D$ and lead to unitary evolution. Again we assume that particle $\AA$ interacts with particle $D$ via a (not necessarily unitary) operation $W_{{\AA}D}$ and require that the final state of particles $D{\BB}$ is related to the initial state of particles $HD$ via a unitary operation $U_{HD\rightarrow D{\BB}}$. We can now write 

\begin{eqnarray}
&& U_{HD\rightarrow D{\BB}}=\<\Phi^+|_{H{\AA}}W_{{\AA}D}|\Phi^+\>_{{\AA}{\BB}}=\nonumber\\
&& =\<\Phi^+|_{H{\AA}}\sum_{ijkl}w_{ijkl}|i\>_{{\AA}}|j\>_{D}\<k|_{{\AA}}\<l|_{D}|\Phi^+\>_{{\AA}{\BB}}=\nonumber\\
&& =\sum_{ijkl}w_{ijkl}|k\>_{{\AA}}|j\>_{D}\<i|_{{\AA}}\<l|_{D}=\nonumber\\
&& =(\sum_{ijkl}w_{ijkl}|i\>_{{\AA}}|j\>_{D}\<k|_{{\AA}}\<l|_{D})^{T_{\AA}}=W^{T_{\AA}}.
\end{eqnarray}

Hence, any operation whose partial transpose is unitary in the three particle picture leads to unitary evolution in one particle picture. In particular, projection onto the maximally entangled state, which we considered in the case of teleportation through polyamorous entanglement, leads to such evolution.

\section{Black hole complementarity as chronology protection}
\label{sec:complementarity}

Let us revisit the notion of black hole complementarity, and see that, in the present context, it can be reinterpreted as a mechanism to prevent violations of causality.

Recall that black hole complementarity was postulated to prevent a violation of cloning by black holes, and that it proves to be insufficient in the case of the AMPS experiment.
In order to witness a violation of the no-cloning theorem in a black hole which is very far from its Page time, an observer outside the black hole
would have to: (i) collect sufficient Hawking radiation (system $\BB$) in order to reconstruct the state $|\Psi|rangle$ which had been thrown into the black hole (system $H$), then (ii) jump into the black hole and catch up with $H$, thereby witnessing that both system $H$ and part of system $\BB$ were in   the same   state $|\Psi|rangle$.
Black hole complementarity postulates that the time it takes to collect enough Hawking radiation to reconstruct $|\Psi|rangle$ on $\BB$ is so long, that by the time
the observer jumps into the black hole, system $H$ has already hit the singularity.

In the case when the system on $\AA$ is not a clone of $H$, but rather the future of $H$, then black hole complementary does not protect us from cloning:  
it   instead   protects us
from having a closed time-like curve. Namely, it  prevents an observer from causing a system $\AA$ to interact with its past at $H$.
We have discussed such chronology violations in Section
\ref{sec:ctc}.
If black hole complementarity prevented
system $\AA$ from interacting with its past at $H$ by delaying its emission, this might prevent a closed time-like curve from being created. However, this does not appear to be an option here.
The reason is that we are already envisioning a scenario akin to that of the AMPS experiment. Namely, in AMPS, after the Page time, each emitted Hawking photon is carrying away
information i.e. is entangled with early radiation outside the black hole (and is also entangled with its
infalling partner). Thus, black hole complementarity is not enough to prevent an observer from witnessing a violation of entanglement monogamy, because the violation occurs for each emitted
photon. Likewise in the scenario considered here, each emitted photon is the future of its infalling partner as it is carrying away information due to the postselection.


\section{Conclusions}
\label{sec:conclusion}

  In conclusion, if we impose unitary evolution on black hole evaporation processes, it appears that while we must weaken the standard causal structure of general relativity to avoid an information paradox, we do not have to violate basic properties of standard quantum theory such as no-cloning and monogamy of entanglement. Of course, if we believe information is 
destroyed in such processes~\cite{bps,unruh-wald-onbps,OR-intrinsic,unruh2012decoherence}, then no paradox exists.  

  In particular,   we have shown that the Horowitz-Maldacena model fits into a temporal product picture.   For example,   even though the presence of 
three particles   in Fig.~1   suggests a tensor product structure, 
the post-selected teleportation   in this model   
transforms the tensor product structure into a temporal product 
of three systems. This is an interesting example where a modification at high energy (the singularity) effects the physics at low energy (at the horizon). Remarkably, it modifies
the physics at the horizon without being inconsistent with effective field theory at the horizon.
We have shown that in the HM case, the violation of monogamy and the no-cloning theorem is 
only apparent: polyamory is allowed in temporal correlations in standard quantum mechanics, 
while cloning does not occur because the obtained ''clones'' are not independent. 

We have   also   shown that the HM model exhibits the same statistics as one obtained from a single 
particle that moves super-luminally at some point. One can pass between both pictures -- the HM model with three particles, 
and the single-particle picture -- by means of Lorentz transformation.
%
%
Superluminal particles are generally considered a highly undesirable feature of a theory, although here we see that it is perhaps less undesirable than the strictly stronger cloning or monogamy violations.
We have also analysed possible interactions in the HM model, and argued that 
the allowed interactions are those that are unitary 
in the single particle picture.  
This   evolution postulate turns out to 
  appears to ensure consistency with  
the process of post-selected teleportation,   including the apparent lack of unitarity that can arise in the latter picture.   


We have also found that there the violation of entanglement monogamy is very different in
the HM proposal compared to the original AMPS formulation. Considering the trio of systems
of early radiation, late time radiation, and infalling radiation, we find that
in the original AMPS picture, only one system is entangled with the other two, while in the
HM proposal, each system is entangled or correlated with the other two. Perhaps this suggests ways out of the AMPS paradox, by adding additional entanglement rather than trying to break it.

Regarding possible paradoxes implied by trying to avoid information loss in black holes, 
we   thus   obtain the following competing pictures if we want to keep unitary evolution:

\bei

\item[(i)] If we assume a tensor product structure between a system which is outside a black hole, and itself when it is inside the black hole, then, unless there is a firewall, both unitarity and causality are violated:
unitarity is violated since there is violation of monogamy of entanglement, which in turns implies violation of causality - as monogamy is implied by no-signaling. 

\item[(ii)] If we assume the Horowitz-Maldacena proposal, we do not have problems with monogamy -- the statistics are
equivalent to those obtained from a particle which is super-luminal at some point. 
Therefore we have violation of causality. Furthermore, there are interactions which are difficult to rule out in a natural way, and lead to
closed time-like curves.

\item[(iii)] If we assume a particle, that is kicked away from the singularity and leaves the black hole, thus travelling super-luminally, we obtain the same statistics as those from the HM model.
  There is perhaps a more natural way to rule out closed time-like curves than in the HM model, however, one needs to make an unnatural identification between infalling radiation and the outgoing particle. On the other hand, in the HM model, the modification of quantum theory is at the singularity, where we anyway expect deviations from quantum theory.

  
\eei

The first assumption has the more unwanted consequences than the other two. The second and third assumptions 
have many of the same unwanted consequences, but it would be surprising if one could preserve both unitarity and the causal structure of the black hole space-time.  Seen in this light,
modifications to quantum theory which allow for destruction of information might be the most conservative solution to the black hole information problem \cite{unruh-wald-onbps,OR-intrinsic,unruh2012decoherence}.





{\it Acknowledgments.}
We thank Karol Bartkiewicz, Charles H. Bennett, Lukasz Czekaj, Daniel Gottesman, Debbie Leung, Seth Lloyd, Rob Meyers, and John Preskill for helpful discussion.  
A.G, M.H and R.H are supported by National Science Centre project Maestro DEC-2011/02/A/ST2/00305, ERC AdG QOLAPS No. 291348, and John Templeton Foundation, Grant No. 56033. The opinions expressed in this publication are those of the authors and do not necessarily reflect 
the views of the John Templeton Foundation.
A.G. acknowledges European Social Fund - Operational Programme ’Human Capital’ - POKL.04.01.01-00-133/09-00 
and University of Cambridge for hospitality. {  MJWH is supported by the ARC Centre of Excellence CE110001027, and thanks the National Quantum Information Center of Gd\'ansk for hospitality. }

 \bibliographystyle{apsrev}

\bibliographystyle{apsrev}


\end{document}